\documentclass[english,aps,prper,reprint,showpacs,titlepage,longbibliography]{revtex4-2}   
\usepackage[utf8]{inputenc}
\usepackage[compatibility=false]{caption} 
\usepackage{booktabs} 
\usepackage{tabularx} 
\usepackage{ragged2e} 
\usepackage{caption} 
\usepackage[T1]{fontenc}	
\usepackage{geometry}
\usepackage{times}
\usepackage{hyperref}  
\hypersetup{colorlinks=true,urlcolor=blue,citecolor=blue,linkcolor=blue} 
\usepackage{array}
\usepackage{enumerate}
\usepackage{amsmath}
\usepackage{amssymb}
\usepackage{tikz}
\usepackage{graphicx}
\usepackage{multirow}
\usepackage{tcolorbox}
\usepackage{ragged2e}
\usepackage{float}
\usepackage[utf8]{inputenc}

\usepackage[normalem]{ulem}
\usepackage{setspace} 
\usepackage{natbib}
\begin{document}
\begin{titlepage}

\title{Investigation of the Inter-Rater Reliability between Large Language Models and Human Raters in Qualitative Analysis}

\author{Nikhil Sanjay Borse}
 \affiliation{Department of Physics and Astronomy, Purdue University, 525 Northwestern Ave, West Lafayette, IN-47907, U.S.A.}

\author{Ravishankar Chatta Subramaniam}
\affiliation{Department of Physics and Astronomy, Purdue University, West Lafayette, IN-47907, U.S.A.} 
  
 \author{N. Sanjay Rebello}
 \affiliation{Dept. of Physics and Astronomy / Dept. of Curriculum \& Instruction, Purdue University, West Lafayette, IN-47907, U.S.A.} 


\begin{abstract}

Qualitative analysis is typically limited to small datasets because it is time-intensive. Moreover, a second human rater is required to ensure reliable findings. Artificial intelligence tools may replace human raters if we demonstrate high reliability compared to human ratings. We investigated the inter-rater reliability of state-of-the-art Large Language Models (LLMs), ChatGPT-4o and ChatGPT-4.5-preview, in rating audio transcripts coded manually. We explored prompts and hyperparameters to optimize model performance. The participants were 14 undergraduate student groups from a university in the midwestern United States who discussed problem-solving strategies for a project. We prompted an LLM to replicate manual coding, and calculated Cohen’s Kappa for inter-rater reliability. After optimizing model hyperparameters and prompts, the results showed substantial agreement (${\kappa}>0.6$) for three themes and moderate agreement on one. Our findings demonstrate the potential of GPT-4o and GPT-4.5 for efficient, scalable qualitative analysis in physics education and identify their limitations in rating domain-general constructs.
  
    \clearpage
\end{abstract}

\maketitle
\end{titlepage}
\maketitle

\section{Introduction \&\ Background}

To prepare for careers in STEM, students need to develop core disciplinary ideas, cross-cutting interdisciplinary concepts, and important engineering and science practices \cite{NGSS2013}. Incorporating Engineering Design (ED) projects in a physics course can help meet these goals, and connect ED with Science Thinking (ST) \cite{honey2014,fischer2014interplay,national2012framework,NGSS2013}. 
Research has shown that physics education should also emphasize ‘ways of thinking’ (WoT) along with problem-solving \cite{dalal2021developing,slavit2019stem,slavit2021student,lien2020well,english2023ways,talanquer2010let}. Several studies have focused on "STEM Ways of Thinking", and on developing theoretical frameworks for segregating and characterizing these WoT \cite{dalal2021developing,slavit2021student,slavit2019stem,lien2020well,english2023ways}. Given the context of our study, WoT refers to the ways in which students think, make decisions, act, and participate in their ED projects \cite{dalal2021developing}. A novel contribution of this study is that it shows the potential of LLMs to identify students' WoT as they participate in ED projects. Observation of student actions, such as peer interactions, can provide insight into their thinking process in a naturalistic setting \cite{luna2018teachers, wilkinson2010developing}. Peer interactions help students explore diverse perspectives, skills, share ideas, and reason  \cite{etkina2014thinking,firetto2023embracing}.   


Qualitative research (QLR) has been important in physics education to understand the nuances of the thought process of students and their problem-solving approaches \cite{denzin1996handbook,erickson1985qualitative}.   
However, a challenge in QLR is the prohibitive time required for human coding or thematic analysis \cite{braun2006using,creswell2016qualitative,morse2015critical, saldana2009introduction}. Software tools like NVivo may have streamlined QLR logistics, but humans still need to analyze the data \cite{houghton2015qualitative,jackson2019qualitative}. Consequently, QLR has a limitation in scalability to large numbers of participants. Most of the work in QLR focuses on the detailed analysis of artifacts of a few participants. The desired coding accuracy in QLR makes reliability a crucial part of the process \cite{uysal2021reliable}. Reliability is generally the extent to which the coding process is free from random errors \cite{turgut2010eugitimde}. Inter-rater reliability (IRR) measures the agreement between multiple raters \cite{aiken2009psychological}. Consequently, raters can code to consensus to ensure that their ratings are reliable \cite{siverling2021initiates}. Although there are several ways to analyze IRR, we use Cohen’s ${\kappa}$ in this study for its simplicity, since we compare thematic coding done by two raters (humans coded to consensus treated as one rater, and the LLM treated as the second rater) with fully overlapping codes \cite{tinsley1975interrater}. 
LLMs 
can potentially revolutionize the efficiency of QLR 
in physics education by scaling up the process for large datasets, 
provided that 
LLM coding is reliable \cite{siiman2023opportunities,katz2023exploring,hitch2024artificial,tabone2023using,zhang2023redefining}.

None of the previous studies that used LLMs for qualitative coding 
investigated how model performance changed by optimizing LLM hyperparameters, such as temperature, which controls randomness of the output, and top-$p$, which controls the number of most probable words sampled. In our study, we address this gap in the literature. Several studies explore how IRR is influenced by prompt engineering, 
albeit in different contexts \cite{dunivin2025scaling,liu2025qualitative,xiao2023supporting}. Prompt engineering improves clarity for the LLM by keeping the prompts short, relevant, and generates clear prototypical examples instead of ambiguous real-world examples, as shown by Dunivin in a sociohistorical context \cite{dunivin2025scaling}. 
In this study, we explored different prompting methods to see whether the IRR of LLMs can be improved in the context of ED projects, integrated in a physics course.

Our primary goal in this study is to compare the qualitative analysis of 
audio transcripts done by human raters and an LLM. 
We address the following research questions. 

RQ1: What is the inter-rater reliability (IRR) between state-of-the-art LLMs such as GPT-4.5 and GPT-4o and expert human raters for coding audio transcripts of students engaged in a group discussion during a lab activity in the context of Engineering Design (ED)?

RQ2: To what extent can the IRR of LLMs such as GPT-4.5 and GPT-4o be improved by (a) prompt engineering and (b) optimizing their hyperparameters through the OpenAI API (Application Programming Interface)?

\section{Methods}

In our study, the student groups in a calculus-based physics laboratory course completed projects in which they recorded their peer interactions to discuss strategies for solving ED challenges. 
A human rater 
recorded the audio transcripts. Two human raters coded the audio transcripts to consensus. LLMs such as ChatGPT-4o (GPT-4o) and ChatGPT-4.5-preview (GPT-4.5) played the role of a rater 
\cite{openai2024gpt4technicalreport,openai2024gpt4o,brown2020language}. 
We then segregated the audio transcripts into text segments, 
and prompted the LLM to classify each
text segment based on whether or not it met at least one of the criteria for a given theme in a framework that we adopted to characterize STEM Ways of Thinking \cite{ravi_perc_2024}. 
The IRR between the LLM and the consensus reached by two human raters was studied using Cohen’s ${\kappa}$ to test the reliability of GPT-4.5 and GPT-4o for qualitative analysis \cite{aiken2009psychological}. 

More specifically, our study occurred in a calculus-based, first-semester undergraduate physics course at a large university in the Midwestern U.S. In Weeks 8-14, students were allowed to choose their own 
ED project.  At the end of 
week 14, 
students were asked to engage in and record a free-flowing discussion for at least five minutes on applying physics and math concepts in their ED projects and how their problem-solving approach evolved over the weeks. For this study, we analyzed data from 14 student groups of three students each. These 14 groups were from one lab section, which was a subset of more than 500 groups enrolled for the course. Consistent with the guidelines for our Institutional Review Board approval, our data were anonymized so that the identity of the participants was not revealed while the analysis was carried out. Based on student responses, four ways of thinking were identified in our framework: Engineering Design (ED), Physics Concepts (PC), Math Constructs (MC), and Metacognitive Thinking (MT) (see Table \ref{tab:ed_example}) \cite{ravi_perc_2024,PhysRevPhysEducRes.21.010118,subramaniam2025applyingstemwaysthinking}.

\begin{figure}
    \centering   \includegraphics[width=1.0\linewidth,height=5.5cm]{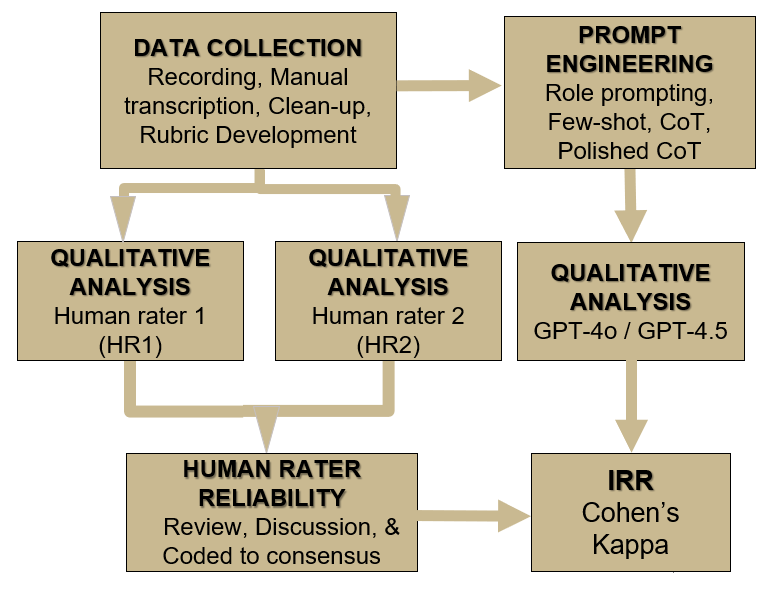}
    \caption{Methodological Flow: Inter-Rater Reliability
    \cite{bijker2024chatgpt}}
    \label{fig:enter-label1}
\end{figure}

Our study combines qualitative analysis by human raters (HR) and LLMs using quantitative methods for IRR. Fig. \ref{fig:enter-label1} shows the methodological flow of our study.
The audio data from the peer interaction was transcribed and manually cleaned. Due to the low audio quality, the transcripts were representative of the group, as it was not feasible to detect them speaking individually. The audio transcripts were qualitatively coded according to the framework or rubric in a previous study \cite{ravi_perc_2024}. 
Each text segment was labeled depending on whether or not it met at least one of the criteria for any given theme in the framework. A text segment could belong to more than one theme from the rubric shown in Table \ref{tab:ed_example}. For the ‘dependability’ and ‘trustworthiness’ of our analysis, we follow Guba and Lincoln \cite{guba1994competing}. Two coders coded the 14 audio transcripts, reviewed, and discussed to consensus \cite{siverling2021initiates}.

\begin{figure}
    \centering \includegraphics[width=1.0\linewidth,height = 5.5cm]{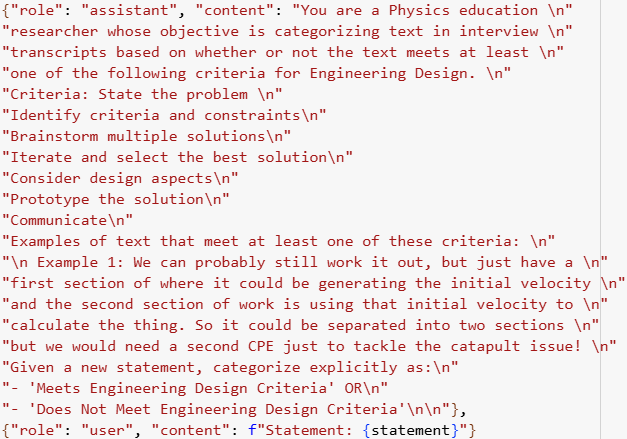}
    \caption{\justifying Example of a polished few-shot prompt.}
    \label{fig:one-shot2}
\end{figure}

Research has shown that LLM responses are sensitive to 
prompt engineering, which was necessary to get reliable ratings using LLMs \cite{mizumoto2025large}. 
Our prompt had instructions about the role of a text classifier (role prompt), criteria for a given theme in the framework with human-labeled example quotes as shown in Table \ref{tab:ed_example} (few-shot prompt), and the text segment to be labeled  \cite{chen2023unleashing,ravi_perc_2024}. Due to the complexity of the task, 
we divided the prompt using triple quotes \cite{chen2023unleashing}. 
We first tested a ‘zero-shot' prompt without any example quotes. After testing on the text segments in three of the 14 transcripts through OpenAI's user interface, it became clear that the few-shot prompt generally yielded better text classification, which aligns with prior studies \cite{liu2025qualitative}. 
The prompt typically had three examples that met the criteria for a given theme, such as ED, and three examples that did not meet any of the criteria. The LLM was tasked with doing a binary classification accordingly for each theme, one text segment at a time, as preliminary tests showed that a decomposed coding approach generally yielded better classification for a single task, instead of classifying many text segments at once \cite{xiao2023supporting}. We then asked GPT-4o to polish the few-shot prompt, which simplified and improved its clarity, and generated prototypical examples that were unambiguous to process for the LLM \cite{dunivin2025scaling}. The polished few-shot prompt resulted in increased reliability of qualitative coding for all themes at a low computational cost \cite{liu2025qualitative}. 
The polished few-shot prompt for the ED theme is shown in Fig. \ref{fig:one-shot2}. The prompts for the other three themes followed the same structure and were polished likewise. 


To find agreement between human raters and LLMs, we performed an IRR by calculating Cohen’s ${\kappa}$ for each theme.
There were 204 text segments 
excluding the example segments in the prompt. To classify them, we used OpenAI's API for batch processing with GPT-4.5 and GPT-4o using 
polished few-shot prompts for decomposed coding. 
We optimized the LLMs for performance. Model hyperparameters like 
temperature and top-$p$, \cite{chen2023unleashing} were fine-tuned. 
The theme descriptions and example quotes are in Table \ref{tab:ed_example}.

\begin{table*}[hbpt]
    \centering
    \caption{\justifying Coding rubric with descriptions and example quotes \cite{ravi_perc_2024}. Engineering Design (ED), Physics Concepts (PC), Math Constructs (MC), and Metacognitive Thinking (MT)}
    \begin{tabular}{p{0.01\linewidth}@{\hspace{0.5cm}}p{0.38\linewidth}p{0.53\linewidth}}
        \hline \hline
        \textbf{Code} & \textbf{\hspace{1cm}Code Description} & \textbf{Example Quote} \\
        \hline
        ED & State the problem; identify criteria and constraints; brainstorm multiple solutions; iterate, select the best solution; consider design aspects; prototype the solution; communicate. & We will focus on the batter’s perspective and calculate the exact time, position, and technique that should hit the ball in order to get the best outcome. We will explore the specific question: What are the optimal conditions for a baseball player to hit a home run?\\
        PC & Identify related physics terms, concepts, or principles; cause and effect; system and surroundings; scale; change and rate of change. & The physics concept was Newton’s II law. We used that so that we’ll know the constant speed over time which means there will be no acceleration.\\
        MC & Mention a formula, equation, or a mathematical concept; refer to a scientific statement 
        of a relation among several variables; proportional reasoning; units analysis; use of explicit equations. & One of the math concepts for this lab was relabeling x and y coordinate vectors or having them in different positions. This is like linear algebra where we rearrange coordinate vectors as basis vectors. \\
        MT & Reflect on their design and science ideas, and progression towards the solution & In our first iteration attempt to solve this problem, we did during lab 11 but this problem did not have... we had too many variables which we didn’t know and it made it too hard to solve this problem.
        \\[0.1cm] \hline\hline
        \end{tabular}
    \label{tab:ed_example}
\end{table*}


\section{Findings \&\ Discussion}
Our primary goal in this study was to compare human coding of 
audio transcripts 
with GPT-4.5 and GPT-4o coding in the context of ED projects, and to compare the performance of the two models. 
To test the reliability of LLM's rating of the transcripts, we did 
an IRR using Cohen’s ${\kappa}$ with two human raters who coded to consensus. For Cohen’s ${\kappa}$, scores between 0.8 to 1.0 were indicative of perfect agreement, 
0.6 to 0.8 were indicative of substantial agreement, 0.4 to 0.6 of moderate agreement, 0.21 to 0.40 of fair agreement, 0 to 0.2 of slight agreement, and below 0 of no agreement 
\cite{landis1977application}. 

\begin{figure}[t]
    \begin{minipage}{1.0\linewidth}
        \centering     \includegraphics[width=\linewidth,height=5cm]{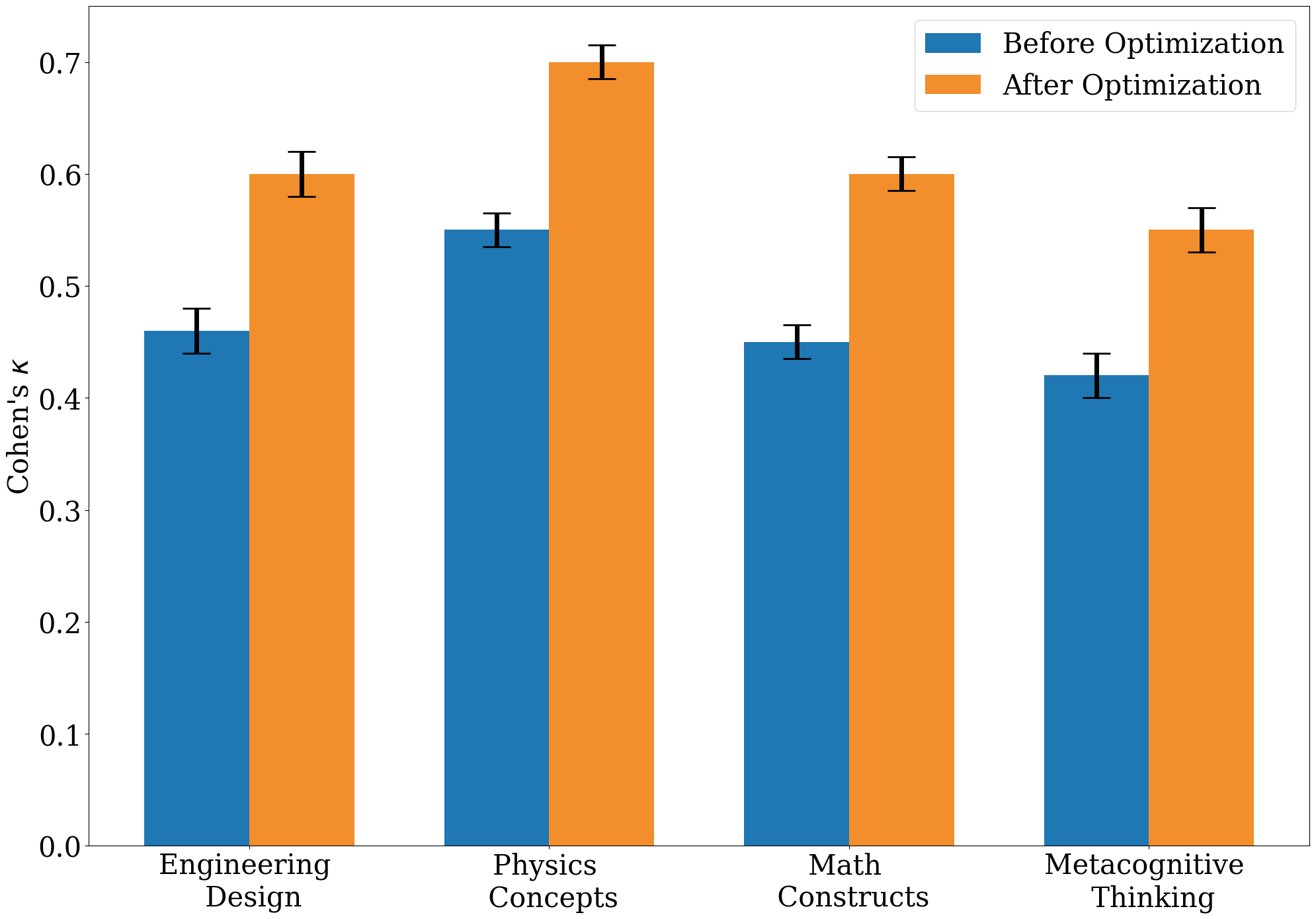} 
        \caption{\justifying Improvement in rater agreement for 
        ED, PC, MC, \&\ MT  using optimal settings, prompts, and models (orange) for each category, as described in Table \ref{tab:code-settings},
        compared to using GPT-4o (blue) with default settings, T = 1 and top-p = 1.}
        \label{fig:rater-improvement}
    \end{minipage}
\end{figure}


The mean values of Cohen’s ${\kappa}$ for 5 runs of each theme are shown in Fig. \ref{fig:rater-improvement}, both with default API settings (blue bars), 
and with optimal settings and polished few-shot prompts (orange bars).
For ED, PC, and MC 
, we found that one of GPT-4.5 or GPT-4o coded them reliably (Cohen’s ${\kappa}$ > 0.6) after optimization \cite{dalal2021developing,slavit2021student,slavit2019stem,lien2020well,english2023ways,dunivin2025scaling}. 
For PC, Cohen's ${\kappa}$ = 0.7, which showed remarkable agreement with human raters. This can be due to the objective clarity of the PC criteria, 
making them easier to rate for both LLMs and human raters \cite{liu2025qualitative}. 
GPT-4o delivered the best results for domain specific themes like PC and MC \cite{xu2024evaluating}. 
GPT-4.5 delivered better results for ED. 

\begin{figure}
    \begin{minipage}{1.0\linewidth}
        \centering     \includegraphics[width=\linewidth,height=5cm]{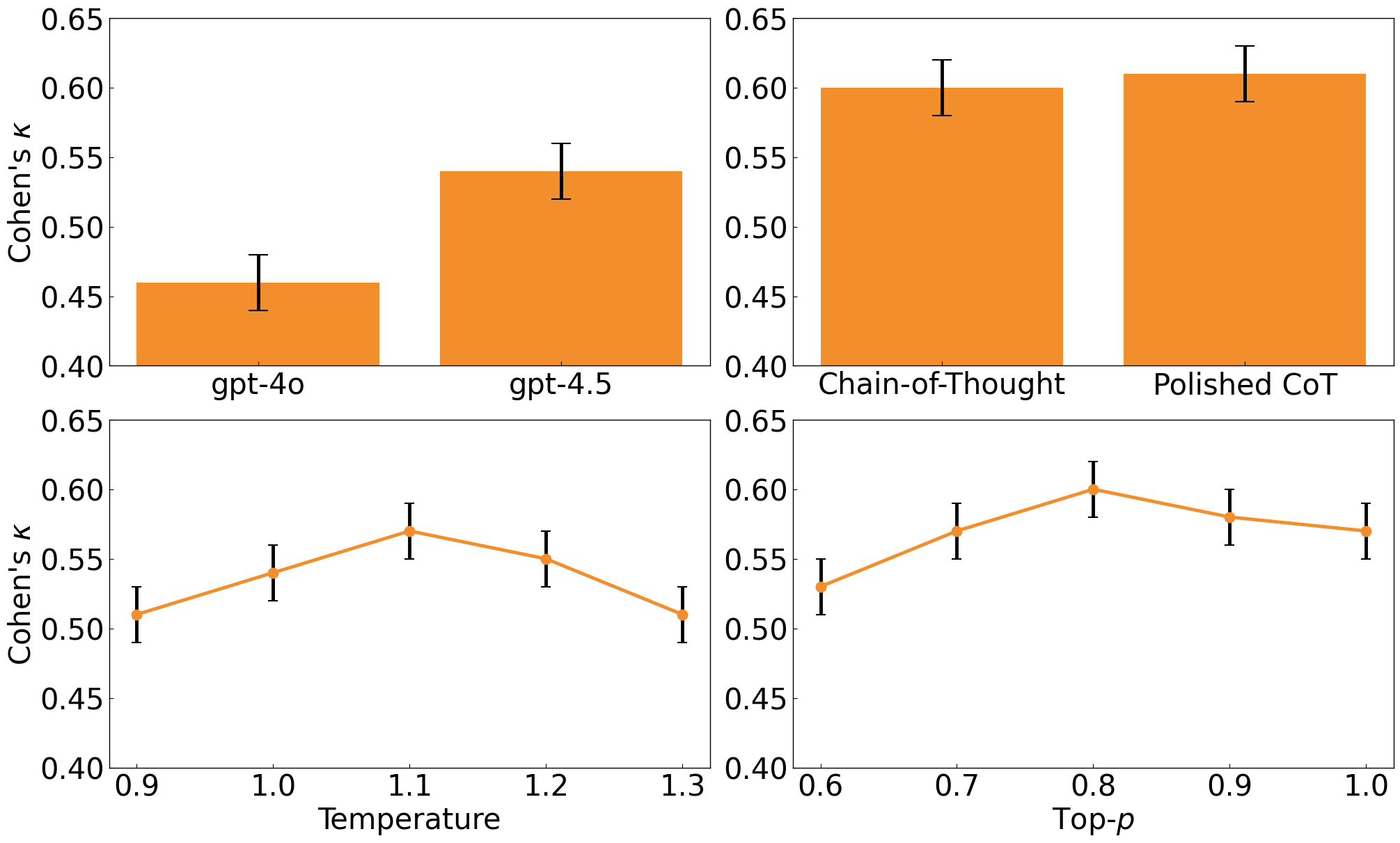}
        \caption{\justifying Cohen's ${\kappa}$ for Engineering Design (ED) by model (top-left), Temperature using GPT-4.5 (bottom left), top-p using GPT-4.5 (bottom right) with T=1.1, and prompting method using GPT-4.5 and top-p = 0.8 (top-right). }
        \label{fig:enter-label}
    \end{minipage}
\end{figure}

Our secondary research goal was to investigate whether (a) optimizing model hyperparameters and (b) prompt engineering methods can improve the performance of LLMs for IRR. We did a detailed analysis for each theme individually or followed a decomposed coding approach to explore (a) optimal hyperparameter settings and (b) prompt combinations \cite{xiao2023supporting}. 

\newcolumntype{Y}{>{\justifying\arraybackslash}X}
\newcolumntype{C}{>{\centering\arraybackslash}X}

\begin{table}[H]
\centering
\caption{\justifying Optimal model settings used with polished few-shot prompts\cite{dalal2021developing,slavit2021student,slavit2019stem,lien2020well,english2023ways}}
\renewcommand{\arraystretch}{1.2}
\begin{tabularx}{\columnwidth}{CCCC}
\hline\hline
Theme & Model & Temp & Top-p \\
\hline
ED & GPT-4.5 & 1.1 & 0.8 \\
MT & GPT-4o   & 1.1 & 0.8 \\
MC & GPT-4o   & 0.9 & 0.9 \\
PC & GPT-4o   & 0.9 & 0.9 \\
\hline
\end{tabularx}
\label{tab:code-settings}
\end{table}

A detailed 
analysis specifically for the 
ED theme can be seen in Fig. \ref{fig:enter-label}. The top left panel of the figure shows how 
GPT-4o and GPT-4.5 performed at rating the text segments with default settings in the API and a few-shot prompt that was not polished. GPT-4.5 was the more reliable model for ED. 
The model selection showed a significant influence on the IRR, which aligns with prior studies that have shown LLMs such as GPT-4o outperform legacy models like GPT-4 \cite{xu2024evaluating}. 
The top-right panel of the figure shows the effect of a polished few-shot prompt on IRR and aligns with prior studies \cite{liu2025qualitative}. 
There is a noticeable improvement in 
Cohen's ${\kappa}$. This works for all themes 
and builds on Dunivin's work, which is in a socio-historical context \cite{dunivin2025scaling}. 

In the bottom panel, we have shown the variation in Cohen's ${\kappa}$ with first the temperature (bottom-left) and then top-$p$ (bottom-right). Both distributions or trendlines show a clear peak at Temperature = 1.1 and top-$p$ = 0.8, respectively. These values of Cohen's  ${\kappa}$ were averaged over 5 runs for each hyperparameter value. 
Higher temperature means more randomness, and 
ED aspects can sometimes be domain general and varied, which could lead to a slightly higher optimal value for temperature than the default value of 1 \cite{chen2023unleashing}. 

Using the combined gains from optimizing GPT-4.5 
and prompt polishing, we achieved
Cohen's ${\kappa}$ = 0.60, an increase of 0.15 (see Fig.\ref{fig:rater-improvement}), which brings us to the borderline between moderate and substantial agreement with human raters \cite{dunivin2025scaling}.

The average increase in Cohen's ${\kappa}$ of all themes was 0.14, which is a statistically significant increase (p < 0.02) using a Mann-Whitney test \cite{mcknight2010mann}.
The optimal settings \&\ prompts for each theme are shown in Table \ref{tab:code-settings}. For 
PC and MC, the optimal settings are 
Temperature = top-$p$ = 0.9, whereas for 
ED and MT, these are Temperature = 1.1, and top-$p$ = 0.8. This makes 
sense as 
ED can be domain general, whereas 
MC and PC are more domain specific, and there is less 
randomness \cite{Jonassen2010LearningTS}. 
Even after the LLMs were used with 
optimal settings 
and polished few-shot prompts, the only theme that showed moderate agreement (${\kappa}$ = 0.55) with human raters was MT. 
This limitation of LLMs can be due to MT not being domain specific, 
which LLMs cannot always reliably rate \cite{liu2025qualitative}. 
Our findings show that LLMs can potentially be used to scale up qualitative analysis to large datasets, while they have limitations in rating domain general constructs.

\section{Conclusion \&\ Implications}
Our first Research Question inquired about the Inter-rater Reliability (IRR) 
between expert human raters and State-of-the-Art LLMs such as GPT-4.5 and GPT-4o in the context of Engineering Design (ED) projects, and performance comparison between the two models. For ED, 
GPT-4.5 showed higher IRR than GPT-4o, and improved agreement with human raters after optimisation. 
We suspect it might be because 
ED is relatively more complex to interpret than Physics Concepts (PC) or Mathematical Constructs (MC), as it can be domain general, and GPT-4.5 is better equipped to process these nuances \cite{openai2024gpt4o}. 
Both models showed 
moderate agreement for Metacognitive Thinking (MT) \cite{xu2024evaluating}. 
This may be 
due to MT not being domain specific \cite{liu2025qualitative}.
GPT-4o showed a higher IRR, and increased agreement with human raters for PC and MC after optimization, as they are domain specific and easier to code both for human raters and LLMs \cite{liu2025qualitative}.

Our second Research Question inquired whether model performance can be improved by optimizing model hyperparameters and prompt engineering. For GPT-4o and GPT-4.5, we compared Cohen's ${\kappa}$ obtained using the default settings in OpenAI's API, with ${\kappa}$ from optimized settings. 
We found a considerable improvement in the 
IRR as the ${\kappa}$ values increased by more than 0.14 for each theme after optimization 
(see Table \ref{tab:code-settings}, and Fig.\ref{fig:rater-improvement}) \cite{dunivin2025scaling,liu2025qualitative}. 
After optimization, 
the agreement between the LLMs and human raters improved significantly across all themes,but despite this gain, the agreement for MT was moderate at best.
\cite{liu2025qualitative}. 

We have shown that State-of-the-Art LLMs, after optimization, can be a reliable tool for qualitative analysis of audio transcripts of student conversations, but have limitations in coding themes that are not domain-specific, such as metacognitive thinking. For STEM researchers, LLMs can be valuable for streamlining the qualitative coding of STEM Ways of Thinking and increasing the speed and reliability of analyzing large datasets \cite{dalal2021developing,slavit2019stem,slavit2021student,lien2020well,english2023ways}. However, 
human-rater oversight is necessary for reliability and ethical rating practices. A small number of human raters can potentially employ and monitor an LLM for qualitative analysis of large datasets. Moreover, the LLMs used in this study through OpenAI's API require a subscription and may not be equitably accessible to everyone. 

This study shows the promise that LLMs like GPT-4.5 and GPT-4o hold for the future of qualitative analysis in physics education. Rapid advances in AI can make qualitative coding faster and reliable for large data sets without sacrificing rigor and nuance. Thematic analysis is of interest to Physics Education Researchers as it can provide vital insights into the richness of students' ways of thinking in various situations  \cite{dalal2021developing,slavit2019stem,slavit2021student,lien2020well,english2023ways}. The Ways of Thinking (WoT) analysis reveals how students think, make decisions, and act in their interdisciplinary ED projects, 
and we might see new themes emerge from a larger dataset \cite{dalal2021developing,etkina2014thinking,firetto2023embracing,talanquer2010let}. The potential emergence of novel themes or WoT could provide pedagogical insights and have implications for scalable personalized feedback, which would also be of interest to STEM educators and researchers. 

\section{Limitations \&\ Future Work}
A major limitation is that we do this reliability study for a small subset of the data. The ways of thinking that emerged from this data 
may not be representative of 
broad population-level trends. Another limitation is the risk that the optimization of models might be overfitting the hyperparameters to our dataset, and may not necessarily generalize well to new data. Future work can use these findings as starting points for generalizibility tests by scaling up the analysis to new large datasets. 
Our study only uses LLMs from OpenAI, 
while there are several other LLMs such as Deepseek-R1 that 
we can explore in future work \cite{guo2025deepseek}. Since OpenAI LLMs are proprietary and costly, the accessibility of these models can be a limiting issue for researchers with severe funding constraints.
Traditional machine learning (ML) has also 
not been investigated here. 
Traditional ML can be employed and tested for qualitative coding and compared with LLMs. Even state of the art LLMs show moderate agreement with human raters for a theme that is not domain specific, like Metacognitive Thinking (MT). Future works might require fine-tuning or training traditional ML models specifically for rating MT, which has thus far been resistant to automation. Unsupervised ML can help expert human raters identify new themes in large datasets based on computational grounded theory \cite{Ai_qlr_tschisgale_2023}. 



\section{Acknowledgments}
This work is supported in part by U.S. National Science Foundation Grant 2300645. Any opinions expressed here belong to the authors and not the Foundation.

\clearpage
\bibliography{references}


\end{document}